\documentstyle[pra,aps,epsf]{revtex}
    \begin{document}
   
\title{Collective dynamics of internal states in a Bose gas}
     \author{M.~\"O.~Oktel$^{\dagger}$, L.~S.~Levitov }

    \address{Department   of  Physics,   Massachusetts   Institute  of
    Technology, Cambridge, MA 02139\\ } \maketitle \today 

    \begin{abstract}
Theory for the Rabi and internal Josephson effects in an interacting 
Bose gas in the cold collision regime is presented. 
By using microscopic transport equation for the density matrix 
the problem is mapped onto a problem of precession 
of two coupled classical spins. In the absence of an external 
excitation field our results agree with the theory
for the density induced frequency shifts in atomic clocks. 
In the presence of the external field, the internal Josephson effect 
takes place in a condensed Bose gas as well as in a 
{\it non-condensed} gas. The crossover from Rabi oscillations 
to the Josephson oscillations as a function of interaction strength 
is studied in detail.
    \end{abstract}

\section{Introduction}
Recent experiments achieving the trapping and Bose condensation of atoms with hyperfine structure \cite{mhj98,hmw98,hme98,sac98,sis98,mss99,smc99} have created a lot of theoretical interest in the behavior of atomic gases with internal degrees of freedom \cite{tlh98,lpb98,yip99,imo99,kue00,hyi00,ued01,kst01}. Most of this interest has focused on how the presence of internal degrees of freedom effect the spatial behavior, such as the hydrodynamical modes or vortices. An equally important question posed by these experiments is how the internal states of the atoms are effected by interactions. 

In a variety of experiments the transitions between internal states are used to  either manipulate or probe cold gases. Fountain atomic clocks \cite{gch93,lgi98,krc89,gll96} use Ramsey spectroscopy technique to measure the energy differences between internal states very precisely, and have found that this energy difference depends on the densities of the different internal states in the gas. Another class of experiments have used Electromagnetically Induced Transparency \cite{morelevel} techniques to excite the atoms into a superposition of internal states by absorbing a `probe' beam and were able reemit the beam after some delay. These experiments measured the change in the `stopped' light beam as the gas evolved in the superposition state \cite{ldb01,pfm01}. Both of these classes of experiments measure the internal dynamics of the atoms in a normal gas. Experiments involving change in the internal state of the atoms have also been carried out on Bose Einstein Condensates (BEC). In the MIT spin polarized Hydrogen experiment \cite{fkw98,kfw98}, BEC was detected by changing the internal states of the atoms from 1S to 2S using a two photon excitation. Two photon transitions have also been successfully employed by the JILA group in a number of  experiments to create topological excitations \cite{mah99a,ahr01} or phase textures\cite{mah99b} in a Rubidium BEC. One can expect many related experiments to be carried out in near future as the study of cold atomic gases is a rapidly expanding field.   

Although all these experiments try to elucidate the internal dynamics of a cold gas, it is useful to classify them into two groups with respect to the strength of the external field that couples to the internal states. In the weak field limit, experiments probe the internal dynamics of a gas of atoms in the absence of an external field. 
For example, in
the atomic clock experiments, although strong fields at the beginning and the end of Ramsey spectroscopy are required, an essentially free evolution of the wavefunction in a superposition of two internal states is measured. In the MIT hydrogen experiment \cite{fkw98}, the number of excited atoms is a small fraction of the total number, and thus all the effects measured are linear in the excitation rate. Thus the external field is only a weak probe of the free dynamics. On the other 
hand, in the strong field limit the internal dynamics is sensitive to the intensity of the external field, and the number of atoms changing their internal state form a large fraction that may reach 100\%. In the JILA experiments, by applying a $\pi$ pulse all the atoms are transferred from one internal state to the other, and duration of the pulse is the inverse of Rabi frequency, showing that the internal dynamics depends on the strength of the external field. It should be also kept in mind that in all these experiments, even in the weak field limit, the external field has large enough number of quanta to be treated as a classical field. Also we will assume throughout that the excitation is carried out by a coherent field. 

The interaction between two atoms generally depends on their internal states \cite{wbz99,leg01}. For the dilute cold gases considered in this work, interactions can be simply characterized by the s-wave scattering lengths $a_{\alpha \beta}$ which depend on the internal states $\alpha$ and $\beta$ of the interacting atoms. Thus a two state problem will have three different scattering lengths. To be in the purely s-wave scattering limit the gas has 
to be cold enough, so that
  \begin{equation}
\label{quantumgascondition}
\frac{a}{\lambda_{T}} \ll 1,
  \end{equation}
where $a$ is the largest of the s-wave scattering lengths in the problem and $\lambda_{T} = h(2 \pi m k_b T)^{-1/2} $ is the thermal de Broglie wavelength. The temperature range defined by the criterion
(\ref{quantumgascondition}) corresponds to the so-called cold collision regime.

It is important to realize that the inequality (\ref{quantumgascondition}) not only defines the condition for the s-wave scattering to be dominant, but also is a criterion for quantum statistical effects (indistinguishibility of the particles) to be observable. 
To see this, consider an evolution of the state of one particular 
atom.
This atom will experience two different kinds of scattering processes: the coherent forward and backward scattering which preserves correlation of the momentum state with the internal state, as well as ordinary elastic collisions causing scattering at an angle different from 0 or $\pi$ which destroy this correlation \cite{bas81,lla82}. The forward and backward processes occur with the `rate'
\begin{equation}
\tau_{coh}^{-1} = \frac{ 4 \pi \hbar a}{m} n,
\end{equation}
 while the ordinary elastic collisions rate is
\begin{equation}
\tau_{coll}^{-1} = 4 \pi a^2 n v_{T},
\end{equation}
where $v_{T} = \sqrt{ 2 k_B T/m}$ is the thermal velocity. Comparing these two rates one can see that the coherent processes happen at a much higher rate in the temperature range (\ref{quantumgascondition}). It is then possible to measure the frequency shifts of the internal states and decide the statistics of the system. The gases satisfying the condition (\ref{quantumgascondition}) are called quantum gases \cite{bas81,lla82,lru84,ole99}.

For the densities relevant for the alkali gas experiments, the temperatures below which the gas becomes quantum in the sense (\ref{quantumgascondition}) are much larger than the Bose-Einstein condensation temperature. [This is true because quantum degeneracy becomes important only at the lower temperatures for which the thermal de Broglie  wavelength is the order  of average interparticle spacing,  $  \lambda_{\rm  T}  \simeq n^{-1/3}$.]
 For  Bose  systems  the substantial  overlap between wavefunctions of different particles and bosonic tendency to be in the same  state then  causes Bose--Einstein condensation,  where a finite fraction of the particles of the gas share the same spatial wavefunction.    The  internal  states   of  the particles  define  the interactions  between  the particles,  thus determine  the self-consistent spatial  wavefunction that  all the particles  share \cite{mhj98,hme98},  transition  between internal states are also effected in return. Most importantly, there are no backward(exchange) scatterings between two bosons sharing the same spatial wavefunction, and the  mean field energy shift for a particle in the condensate is different from a particle in the normal gas.

An important example of the modification of the transitions between the internal states upon condensation is given by the internal Josephson effect \cite{wwc99,ost99,vle99,rwa97,helium3,leg01}. In this effect, the internal states act as the two reservoirs and the external field provides the weak link between these two reservoirs 
similar to the tunneling coupling through a barrier in the ordinary spatial Josephson effect \cite{jav86,zsl98,rsf99,sfg97}. Although the JILA experiments \cite{hmw98,hme98} have come close, the internal Josephson effect has not been observed in the alkali gases to date, mostly due to the fact that the shape of the condensate is  different for the two internal states. Any transition between the two internal states causes spatial disturbances, making the internal Josephson oscillations hard to observe. It should however be possible observe the effect by either using a very shallow trap or going to the adiabatic limit in which the Rabi frequency for transitions is much larger than the oscillation frequencies of the condensate in the trap \cite{wwc99,leg01}. We will consider the case of a homogeneous gas and investigate the Internal Josephson effect, having in mind a gas sample in a very shallow trap.

To understand these different phenomena in a general framework, we first derive the transport equation for the  density matrix of the system, treating the  external  degrees of  freedom  semi--classically.  The  transport equation   can  be  used   to  understand transitions
between internal states of a Bose gas in the presence of  spatial non-uniformity caused by a trapping  potential, as well as in the presence of macroscopic  flows, such  as  currents  created by  a  vortex.

Below we focus on the dynamics of internal states in a spatially uniform system. We assume that all the  excitation fields are applied uniformly on a homogeneous sample.  We find the  equations of motion for the internal state density matrix of both the normal component and the condensed  part of a Bose gas. Then we  use a Bloch sphere representation  \cite{API}  for  the  density  matrix  to  get  an equivalent equation of motion for classical spins.

After  deriving the  equations of  motion we  consider  two simple cases: a normal gas and a fully condensed gas. We find the free precession frequencies for the relative phase of two internal states, which leads to an expression for the density dependence of the frequency shift in  fountain atomic clocks. Our results for the  normal gas agree  with previous  work \cite{kvg97,vks87,tvs92}, and  we present  a new result for the fully condensed case.

Then we  investigate the response of  the Bose gas  to an external field when it is either  above condensation temperature or at zero temperature, {\it i.e.} fully  condensed. For the fully condensed Bose gas we show how the Rabi oscillations in a non-interacting gas turn into the internal Josephson oscillations, and calculate the oscillation frequencies exactly for the entire parameter range. By repeating the same analysis for a non-condensed gas, we find that
the internal Josephson effect exists in a cold non-condensed gas. This somewhat surprising result shows that the Josephson effect does not require a broken symmetry, and can be observed in any coherent system. We calculate the frequencies for the normal case and compare with the results for condensed gas. We also compare the internal Josephson oscillations with the spatial Josephson oscillations discussed in \cite{jav86,zsl98,rsf99,sfg97} and show that `macroscopic quantum self trapping' effect is also present for the internal Josephson effect.

In the next section, we study  the internal dynamics of a Bose gas which has comparable densities of condensed and normal components. We first discuss the dynamics without an external coupling field and calculate the oscillation frequencies for the condensate density in each internal state. In the limit of small oscillation amplitude out result agrees with the resonance frequencies calculated for the MIT hydrogen spectrum \cite{ole99}. We briefly discuss the behavior of the system under a strong external field.

Finally we give the transport equation and the equations of motion for the internal  density matrix of a Fermi  gas. We consider a freely  precessing state and its response to an external field, and compare the results with those for the Bose gas.

\section{The  transport  equation}  A   dilute  Bose  gas in a confining potential $U(r)$ can  be described by the Hamiltonian
\begin{eqnarray}   
\label{spatialH}  
{\cal   H}   =  &\int&   d^3r \sum_{\alpha} \psi^+_{\alpha}(r) \left( -\frac{\hbar^2 \nabla^2}{2 m} +  \hbar U_{\alpha}(r) \right) \psi_{\alpha}(r)  \\ \nonumber 
+ &\int&  d^3r   \sum_{\alpha,\beta}  \frac{  \hbar  \lambda_{\alpha \beta}}{2}  \psi^+_{\alpha}(r)  \psi^+_{\beta}(r)  \psi_{\beta}(r) \psi_{\alpha}(r),    
\end{eqnarray}  
where   the   Greek  indices $\alpha$,  $\beta$   run over  all internal states.   The potential  energy  $\hbar  U_{\alpha}(r)$  includes  the internal energies $\hbar  w_{\alpha}$ of the states $|\alpha\rangle$, $\hbar  U_{\alpha}(r)=\hbar  U(r) + \hbar  w_{\alpha}$, and the interaction parameters   $\lambda_{\alpha  \beta}$   are  related   to  the s-wave scattering   lengths   $a_{\alpha   \beta}$  as   
\begin{equation}
\lambda_{\alpha \beta}  = \frac{4 \pi  \hbar}{m} a_{\alpha \beta}.
\end{equation}  
For a  problem with $N$ internal levels there are $ N  (N+1)/2  $  interaction strengths $\lambda_{\alpha \beta}$ to be specified.  The operators  $\psi(r)$ are  the canonical second-quantized Bose operators, satisfying 
  \begin{equation}
\label{Bosecom}       
[\psi_{\alpha}(r),\psi^+_{\beta}(r')] = \delta_{\alpha \beta} \delta(r-r').  
  \end{equation}
Transitions  between internal  states caused by an excitation  field, 
are described by another  term to  be added  to the Hamiltonian (\ref{spatialH}):         
\begin{equation}
\label{excitation}  {\cal  H}_{\rm  exc}=\sum_{\alpha \beta}  \int d^3r  V_{\alpha  \beta}(r,t)  \psi_{\alpha}^+(r)  \psi_{\beta}(r).
\end{equation} 
We assume harmonic time dependence  
\begin{equation} 
V_{\alpha  \beta}(r,t) = A_{\alpha   \beta}(r)    \exp   (i   \Omega_{\alpha    \beta}   t)
\end{equation}  
where $A_{\alpha \beta}(r)$ is a function of coordinates 
describing spatial distribution of the excitation field \cite{morelevel}.  The requirement of  being hermitian gives $V_{\alpha \beta}(t) = V^*_{\beta \alpha}(t)$.

In the case of Bose condensation we can treat the $\psi$ operators as having non-zero expectation values. This is achieved by a substitution
\begin{equation}         
\hat{\psi_{\alpha}}(r)  \rightarrow \hat{\psi}_{\alpha}(r)  +  \bar{\psi}_{\alpha}(r),  
\end{equation}
where $\bar{\psi}_{\alpha}$ is a c-number representing the condensate wavefunction.

To describe the dynamics of the internal states of the normal gas, we need the populations of each level, as well as the coherences between  any two  states. Thus we  define the  full density  matrix $\varrho(r,p)$ and the internal  density matrix $\rho(r)$, both being $N\times N$  hermitian   matrices  for  $N$  internal  states:
 \begin{eqnarray}  
\label{normalmatrix} 
\varrho_{\alpha \beta}(r,p) =  \int  d^3r' \langle \psi^+_{\alpha}(r+\frac{r'}{2}) \psi_{\beta}(r-\frac{r'}{2}) \rangle  e^{i \vec{p} \cdot \vec{r}'} \\ \nonumber 
\rho_{\alpha  \beta}(r) = \int \frac{d^3p}{(2 \pi)^3} \varrho_{\alpha \beta}(r,p) =  \langle \psi^+_{\alpha}(r) \psi_{\beta}(r) \rangle.  
\end{eqnarray}

The internal dynamics of  the condensate can be described by specifying complex amplitudes of internal states, in total $N$ complex  numbers for an $N$ component condensate. We find it useful, in order to keep track of relative phases of condensate internal states, to introduce an $N\times N$ matrix similar  to the density  matrix of normal  states $\bar{\rho}_{\alpha \beta} (r)$. We will   also need  to  keep   track  of   the   condensate  flow described by $\vec{J}_{\alpha \beta}$. 
These quantities are defined as
\begin{eqnarray} 
\label{condansatematrix}
 \bar{\rho}_{\alpha \beta} (r)  &=&  \bar{\psi}_{\alpha}^*(r) \bar{\psi}_{\beta}(r) \\ \nonumber  
\vec{J}_{\alpha \beta} (r) &=& i  \frac{\hbar}{2   m}  \left[  (\nabla  \bar{\psi}_{\alpha}^*(r)) \bar{\psi}_{\beta}(r) - \bar{\psi}_{\alpha}^*(r) ( \nabla \bar{\psi}_{\beta}(r)  ) \right].   
\end{eqnarray} 
Here one notes that $\bar{\rho}$  is always  rank one, thus the matrix does not contain any additional information but is introduced just for convenience.   The   diagonal   elements  of   $\rho_{\alpha \alpha}(r)$ and  $\bar{\rho}_{\alpha \alpha}(r)$ give  the thermal and condensate populations $n_{{\rm t} \alpha}(r)$ and $n_{{\rm c} \alpha}(r)$,  respectively.  Total   density  of  atoms  in  state  $|\alpha\rangle$  is  $n_{\alpha}(r)=n_{{\rm t}  \alpha}(r)  + n_{{\rm  c} \alpha}(r)$.

We start with deriving  the  equations  of  motion   for  $\varrho$  and $\bar{\rho}$. Using  the  Heisenberg equations  of motion  for  the  field operators \begin{eqnarray}  
\frac{d}{dt} \psi_{\alpha}(r) &=& i [{\cal H}, \psi_{\alpha}(r)] , \\ \nonumber
\frac{d}{dt}  \bar{\psi}_{\alpha}(r)   &=&  i  \frac{\partial{\cal H}}{\partial  \bar{\psi}^*_{\alpha}(r)}   
\end{eqnarray}
and then taking the required expectation values we obtain $\frac{d \varrho}{dt}(r,p)$   and  $   \frac{d \bar{\rho}}{dt}(r)$. The contribution of each  term in the Hamiltonian to  the evolution of $\varrho$ and $\bar{\rho}$ can be evaluated separately. To remind the  reader how  this calculation  is carried  out we  present the contribution      of     potential     energy      to     $\frac{d \varrho}{dt}(r,p)$.    First    we   calculate    
\begin{eqnarray}
\dot{\psi}_{\alpha}(r)  &=&  i  \left[  \int  d^3r'  \sum_{\gamma} U_{\gamma}(r')  \psi^+_{\gamma}(r') \psi_{\gamma}(r') , \psi_{\alpha}(r)   \right]   \\  \nonumber
&=& i \int   d^3r' \sum_{\gamma} U_{\gamma}(r') [\psi^+_{\gamma}(r') , \psi_{\alpha}(r)]  \psi^+_{\gamma}(r')   \\  \nonumber  
&=& - i U_{\alpha}(r)   \psi_{\alpha}(r).
\end{eqnarray}   
Upon  complex conjugation we  get 
\begin{equation} 
\dot{\psi}^+_{\alpha}(r)  = i U_{\alpha}(r) \psi^+_{\alpha}(r).  
\end{equation} 
Then we write the    contribution    of    potential    energy    to    $\frac{d \varrho}{dt}(r,p)$    as    
\begin{eqnarray}    
\dot{\varrho}^{\rm (Pot)}_{\alpha \beta}(r,p) &=& \int d^3r' e^{i p \cdot r'} \left( \langle \dot{\psi}^+_{\alpha}(r+r'/2) \psi_{\beta}(r-r'/2) \rangle + \langle \psi^+_{\alpha}(r+r'/2)  \dot{\psi}_{\alpha}(r-r'/2) \rangle \right)  \\ \nonumber 
&\simeq&  i \int d^3r' e^{i  p \cdot r'} \left(  U_{\alpha}(r) - U_{\beta}(r)  + \vec{r}'\cdot \nabla_r (\frac{U_{\alpha}(r)+U_{\beta}(r)}{2}) \right) \langle \psi^+_{\alpha}(r+r'/2) \psi_{\beta}(r-r'/2)  \rangle \\ \nonumber
&=&  i (U_{\alpha}(r)-U_{\beta}(r)) \varrho_{\alpha  \beta}(r,p) + \nabla_r   (\frac{U_{\alpha}(r)+U_{\beta}(r)}{2})  \cdot  \nabla_p \varrho_{\alpha \beta}(r,p).  
\end{eqnarray}

Going  through  this  procedure  for  each of  the  terms  in  the Hamiltonian we  get the equation  of motion.  Whenever we  need to calculate the  expected value  of a 4  particle operator we  do it using   Wick's  theorem,  
\begin{equation}   
\label{wick}  
\langle \psi^+_{\alpha} \psi^+_{\beta}  \psi_{\beta} \psi_{\alpha} \rangle =   \langle    \psi^+_{\alpha}   \psi_{\alpha}   \rangle   \langle \psi^+_{\beta}  \psi_{\beta}  \rangle  +  \langle  \psi^+_{\alpha} \psi_{\beta} \rangle \langle \psi^+_{\beta} \psi_{\alpha} \rangle. 
\end{equation} 
In this averaging the  first term is referred to as the direct (Hartree) term as it depends only on densities while the  second is the exchange  (Fock)  term that involves coherences.

We  find the  following equations  describing the  space  and time variation of the  two matrices: 
\begin{eqnarray} 
\label{transport}
\lefteqn{\left(\partial_{\rm  t} + \frac{\vec{p}}{m} \cdot \nabla_r -  \nabla_r( \frac{U_{\alpha}(r)+U_{\beta}(r)}{2} ) \cdot \nabla_p \right) \varrho_{\alpha \beta}  (r,p) =} \\ \nonumber 
&+& i \left[ U_{\alpha}(r) - U_{\beta}(r) + \sum_{\gamma} (\lambda_{\gamma  \alpha}   -  \lambda_{\gamma  \beta})  \rho^{\rm tot}_{\gamma  \gamma}(r) \right]  \varrho_{\alpha \beta}  (r,p) \\ \nonumber  
&+& i  \sum_{\gamma} \lambda_{\gamma  \alpha} \rho^{\rm tot}_{\alpha   \gamma}(r)  \varrho_{\gamma   \beta} (r,p) - i \sum_{\gamma}   \lambda_{\beta   \gamma} \rho^{\rm   tot}_{\gamma \beta}(r)   \varrho_{\alpha  \gamma}   (r,p) \\  \nonumber   
&+& \sum_{\gamma}   \frac{\lambda_{\gamma  \alpha}   +  \lambda_{\beta \gamma}}{2}  \nabla_r   \rho^{\rm  tot}_{\gamma  \gamma}(r)  \cdot \nabla_p   \varrho_{\alpha   \beta}   (r,p)   \\   \nonumber   &+& \sum_{\gamma} \frac{\lambda_{\gamma \alpha}}{2} \nabla_r \rho^{\rm tot}_{\alpha  \gamma}(r)  \cdot  \nabla_p  \varrho_{\gamma  \beta} (r,p)  - \sum_{\gamma}  \frac{\lambda_{\beta  \gamma}}{2} \nabla_r \rho^{\rm  tot}_{\gamma \beta}(r)  \cdot  \nabla_p \varrho_{\alpha \gamma} (r,p)  \\ \nonumber  
&+& i \sum_{\gamma}  \left[ V_{\gamma \alpha}(r,t)  \varrho_{\gamma \beta}(r,p) - V_{\beta \gamma}(r,t) \varrho_{\alpha    \gamma}(r,p)   \right] \\ \nonumber
&+& \frac{1}{2}\sum_{\gamma}  \left[  \nabla_r V_{\gamma  \alpha}(r,t) \cdot  \nabla_p \varrho_{\gamma  \beta}(r,p)  - \nabla_r  V_{\beta \gamma}(r,t)  \cdot \nabla_p \varrho_{\alpha  \gamma}(r,p) \right]
\end{eqnarray}     
and     
\begin{eqnarray}     
\label{continuity}
\lefteqn{\left(  \partial_{\rm t}  \bar{\rho}_{\alpha  \beta}(r) + \nabla_r \cdot  \vec{J}_{\alpha \beta}(r) \right)=  } \\ \nonumber
&+& i \left[ U_{\alpha}(r) - U_{\beta}(r) + \sum_{\gamma} (\lambda_{\gamma  \alpha}   -  \lambda_{\gamma  \beta})  \rho^{\rm tot}_{\gamma \gamma}(r)  \right] \bar{\rho}_{\alpha \beta}  (r) \\ \nonumber  
&+&  i  \left[  \sum_{\gamma}  \lambda_{\gamma  \alpha} \rho_{\alpha    \gamma}(r)    \bar{\rho}_{\gamma    \beta}(r) - \sum_{\gamma}   \lambda_{\beta   \gamma}  \rho_{\gamma   \beta}(r) \bar{\rho}_{\alpha   \gamma}(r)  \right]   \\   \nonumber  
&+&  i \sum_{\gamma}  \left[  V_{\gamma  \alpha}(r,t)  \bar{\rho}_{\gamma \beta}(r)  - V_{\beta  \gamma}(r,t)  \bar{\rho}_{\alpha \gamma}(r)  \right]     
\end{eqnarray}    
where     $\rho^{tot}(r)=    \rho(r)
    +\bar{\rho}(r)$.

Let us summarize here the approximations used to derive the above equations and assess their validity. The Hartree-Fock approximation is used to treat interactions, that is the same as keeping the lowest order term in the 
expansion in the gas parameter $n a^3$. This is an excellent approximation for the dilute atomic gases where $n a^3$ is usually about $10^{-6}$.

The most important restriction on the above equations (\ref{transport},\ref{continuity}) is that they are valid for times not longer than the elastic scattering time 
  \begin{equation}
\tau_{coll}^{-1} = 4 \pi a^2 n v_{T}.
  \end{equation}
However, for a cold gas the characteristic frequency of the evolution described by (\ref{transport},\ref{continuity}) is much higher than $\tau_{coll}^{-1}$, and this approximation gets better as the temperature is decreased. 
The decoherence due to elastic collisions can be described heuristically
by adding a damping term
  \begin{equation}
\dot{\varrho}^{\rm (Damp)}_{\alpha \beta} = - 4 \pi a^2_{\alpha \beta} v_{\rm T} \varrho_{\alpha \beta} (1 - \delta_{\alpha \beta}),
  \end{equation}
on the right hand side of Eq.(\ref{transport}). 

All other effects, such as the difference of the trap potential for 
different internal states, or density oscillations created by a change in the internal state are adequately described by (\ref{transport},\ref{continuity}) within the Hartree-Fock approximation. 
    
\section{Equation of  motion for the internal  density matrix} Our main  interest   in  this  paper  will  be   the  consequences  of interparticle interactions  in the internal state  dynamics, so we will simplify the above general  equations by assuming that we are in  a uniform  system, with  no  normal or  condensate flows,  and further assume that the  external fields are applied uniformly. In this  case  we will  have  no need  for  the  full density  matrix $\varrho(r,p)$,  as  internal  dynamics  is  completely  described $\rho(r)$ which will be the same at all points r.

Thus    equations   of    motion   reduce    to   
\begin{eqnarray}
\dot{\rho}_{\gamma \gamma'} = &i& (w_{\gamma}-w_{\gamma'}+U_{\gamma}-U_{\gamma'})  \rho_{\gamma \gamma'} +  i \sum_{\alpha} \lambda_{\alpha  \gamma} (\rho_{\gamma  \alpha}  + \bar{\rho}_{\gamma  \alpha})  \rho_{\alpha \gamma'}  \\ \nonumber    
-&i&  \sum_{\alpha} \lambda_{\alpha \gamma'} (\rho_{\alpha \gamma'} + \bar{\rho}_{\alpha \gamma'}) \rho_{\gamma \alpha}  +  i \sum_{\alpha}  (  V_{\alpha \gamma}(t)  \rho_{\alpha \gamma'} -  V_{\gamma' \alpha}(t) \rho_{\gamma  \alpha}) \nonumber
\end{eqnarray}  
\begin{eqnarray} 
\dot{\bar{\rho}}_{\gamma \gamma'} = &i& (w_{\gamma}-w_{\gamma'}+U_{\gamma}-U_{\gamma'}) \bar{\rho}_{\gamma  \gamma'}  +  i  \sum_{\alpha}  \lambda_{\alpha \gamma}   \rho_{\gamma  \alpha}  \bar{\rho}_{\alpha   \gamma'}  \\ \nonumber 
-&i& \sum_{\alpha} \lambda_{\alpha \gamma'} \rho_{\alpha \gamma'} \bar{\rho}_{\gamma \alpha}  + i \sum_{\alpha} ( V_{\alpha \gamma}(t)  \bar{\rho}_{\alpha  \gamma'}  - V_{\gamma'  \alpha}(t) \bar{\rho}_{\gamma   \alpha})   
\end{eqnarray}   
where $U_{\gamma}$   are  defined   as,   
\begin{equation}  
U_{\alpha}= \sum_{\beta} \lambda_{\alpha  \beta} (\rho_{\beta \beta}+\bar{\rho}_{\beta    \beta}) =    \sum_{\beta} \lambda_{\alpha \beta} n_{\beta}.  
\end{equation}

The  dynamics   described  in  the  above   equations  are  fairly complicated if there are fields  coupling more then two states, so we will concentrate on the  simplest case, where only two internal states are  coupled, and all  the populations in the  other states remain  constant. In  this  case  we will  be  concerned with  the dynamics of two,  two by two matrices. One for  the normal gas and one for  the condensate.  Let  us assume that  states 1 and  2 are coupled. We  can go to  a Larmor basis  with the frequency  of the coupling field  so that  the elements of  our density  matrix (and similarly the  condensate matrix) can be  redefined.  The diagonal elements  of the  matrices do  not change  while  the off-diagonal elements  change  as 
\begin{equation}  
\rho'_{1  2}  = \rho_{1  2} \exp[-i  \Omega_{12}  t],  \hspace{1cm}  \rho'_{2 1}  =  (\rho'_{1 2})^*.  
\end{equation}

In this basis  we will find it useful to  rewrite the equations of motion  in the  Bloch  representation \cite{API}.   Being  2 by  2 hermitian matrices  we can expand,  
\begin{eqnarray} 
\rho'_{\gamma \gamma'}  &=&  \rho_{0} \delta_{\gamma  \gamma'}  + \vec{S}  \cdot \vec{\sigma}_{\gamma  \gamma'}  \\  \nonumber  
\bar{\rho'}_{\gamma \gamma'}  &=& \bar{\rho}_{0}  \delta_{\gamma \gamma'} + \vec{S}_c \cdot  \vec{\sigma}_{\gamma   \gamma'}    
\end{eqnarray}   
where $\vec{\sigma}$ are the Pauli matrices.

In  this  representation   $\bar{\rho}_0$  and  $\rho_0$  will  be proportional to the  total number of atoms in  the condensate, and above the condensate respectively.  The $z$ components of both spins are  proportional to  the  population difference  between the  two internal states, belonging to the normal gas or the condensate. For the normal part, the norm of the projection of spin onto the $x-y$ plane represents  the  degree  of  coherence between  the  two  internal states.   For the condensate  spin however,  this norm  is directly proportional to the  geometric mean of the populations  of the two internal states. The angle  corresponding to a rotation around the $z$  axis is  related  to the  relative  phase of  the two  internal states, in both cases.

The matrix $V_{\alpha \beta}$  has no diagonal elements as defined in the  excitation Hamiltonian. However, we can  take the detunings that come  as a result  of going to  Larmor basis, as  the diagonal elements. $V_{11} (V_{22})$ being defined as  $ +(-) [ w_2 - w_1 - \Omega_{12}]$. Then we can again  expand $ V_{\gamma \gamma'} $ as 
\begin{eqnarray}  
V_{\gamma   \gamma'}  &=&  V_{0}  \delta_{\gamma \gamma'} + \vec{V} \cdot \vec{\sigma}_{\gamma'  \gamma}.
\end{eqnarray}

In this  representation we can  write the equations of  motion for the  condensate and  the normal  gas spins  as  : 
\begin{eqnarray}
\label{spineom}  
\dot{\vec{S}} &=&  \vec{S} \times  \vec{B}_n  + 2 \lambda_{1 2} \vec{S} \times  \vec{S}_c + 2 \vec{S} \times \vec{V} \\ \nonumber  
\dot{\vec{S}_c} &=&  \vec{S}_c \times \vec{B}_c  + 2 \lambda_{1  2}  \vec{S}_c  \times  \vec{S} +  2  \vec{S}_c  \times \vec{V}  
\end{eqnarray}  
with, 
\begin{eqnarray}  
\label{magfields}
\vec{B}_n  &=& [  (\lambda_{1  1}  - \lambda_{2  2})  (2 \rho_0  + \bar{\rho}_0)  +  (\lambda_{11}+\lambda_{22}-  2 \lambda_{12})  (2 \vec{S} + \vec{S}_c) \cdot \hat{z}] \hat{z} \\ \nonumber 
\vec{B}_c &=& [  (\lambda_{1 1} -  \lambda_{2 2}) (\rho_0 +  \bar{\rho}_0) + (\lambda_{11}+\lambda_{22}- 2  \lambda_{12}) (\vec{S} + \vec{S}_c) \cdot \hat{z}] \hat{z}. 
\end{eqnarray}

It is important  to observe that the equations  conserve the total densities  in the  condensed and  non condensed  fractions  of the gas. To get a better sense  of the equations we can note that they can be derived from the XXZ self interacting Hamiltonian:
\begin{eqnarray}  
\label{spinH}   
\cal{H}  =&  &   (\lambda_{11} -\lambda_{22})  (\rho_0+\bar{\rho}_0)  \hat{z}  \cdot  (\vec{S} + \vec{S}_c)  + ( \lambda_{11}  -\lambda_{22}) \rho_0  \hat{z} \cdot \vec{S} \\  \nonumber 
&  & + J^{ij}  S^i S^j +  \frac{1}{2} J^{ij} S^i_c S^j_c  + J^{ij} S^i_c  S^j \\ \nonumber  
& & + 2  (\vec{S} + \vec{S}_c ) . \vec{V} 
\end{eqnarray} 
where,
\begin{equation} 
J = \left(  \begin{array}{ccc} 2 \lambda_{12} & 0  &   0   \\   0   &   2   \lambda_{12}   &  0   \\   0   &   0   & (\lambda_{11}+\lambda_{22})  \end{array}  \right)  
\end{equation}  
using the Poisson  spin algebra: 
\begin{eqnarray} 
\{S^i,S^j\} &=& \epsilon^{i j k} S^k \\ \nonumber \{S^i_c,S^j_c\}
 &=& \epsilon^{i j k} S^k_c \\ \nonumber \{S^i,S^j_c\} &=& 0.  
\end{eqnarray}

Let us briefly remind what the physical bases for various terms in the Hamiltonian are. The factor of $2$ difference in the normal spin--normal spin interaction compared to the condensate spin--condensate spin interaction stems from the fact that there are no exchange scatterings within the condensate. The terms coupling the condensate spin with the normal spin arise from processes in which an atom in one internal state goes through an exchange scattering process with a condensate atom in another internal state. 

\section{Density shifts in atomic clocks: free precession of a single spin}
We   start   the   analysis   of   the   dynamics   described   in Eq.(\ref{spineom}) by  considering the  situation in which  we can represent  the  system by  a  single  spin.   There are  two  such cases. The first is at zero temperature, when almost all the atoms are in  the condensate; while  the second is above  the transition temperature, when  there is  no condensate present.   However, our analysis is still restricted to temperatures low enough to satisfy
    the quantum gas condition, $\lambda_{\rm T} > a_s$.

We  first analyze  equations of  motion Eq.(\ref{spineom})  in the case where there is no  external field. Such an analysis is needed to understand any interaction related effects in experiments where particles spend  a substantial amount  of time in  a superposition state, such as atomic clocks \cite{gch93,lgi98,krc89,gll96}.

Let  us   first  consider  the   case  where  there  is   no  Bose condensation. Thus  our equations (\ref{spineom})  will be reduced to   
\begin{equation}   
\vec{S}   =  \vec{S}   \times   \vec{B}_n.
\end{equation}

We  can easily  see  that the  z  component of  the  spin will  be conserved and we will only have a precession around the z axis and the     precession    frequency    will     be    
\begin{equation}
\label{densshift}  
\omega= (w_2  - w_1)  + 2  (\lambda_{22}  n_2 - \lambda_{11}    n_1    +    \lambda_{12}    (n_1   -    n_2)) + \sum_{\gamma\not=1,2}  (\lambda_{2 \gamma}  -  \lambda_{1 \gamma}) n_{\gamma}.  
\end{equation} 
This result  agrees with the theory of frequency shifts in atomic clocks \cite{kvg97,vks87,tvs92}.

We  can now  ask the  same question  for a  sample which  is fully condensed. When all the atoms are in the condensate, they will all be sharing the  same spatial wavefunction, this will  mean that no exchange  processes will  be  possible. That  would eliminate  the factor of 2 multiplying the combination of $\lambda$ and $n$'s for states one and two in  the previous equation.  Not surprisingly we get, for a fully  condensed sample 
\begin{equation} 
\omega= (w_2 - w_1) + (\lambda_{22} n_2 -  \lambda_{11} n_1 + \lambda_{12} (n_1 - n_2))  + \sum_{\gamma\not=1,2}  (\lambda_{2  \gamma} -  \lambda_{1 \gamma}) n_{\gamma}.  
\end{equation}

Although it is not desirable to use a condensate in a fountain atomic clock due to very high density shifts that would result. We see here that such an experiment can be used to probe the correlations in a condensate by simply looking at the density induced frequency shifts.

\section{Internal Josephson effect: single spin under an external field}

In this section we consider the response to an external field, only when the  system can be represented  by a single  spin, {\it i.e.} the system is  either fully condensed or not  condensed at all. In the   former   case, when the system is at zero temperature we have two condensates in internal states $1$ and $2$, which are connected by the ``weak link'' supplied by the external field. This is the well studied problem of the internal Josephson effect \cite{wwc99,ost99,vle99,rwa97,helium3}.

The important difference between the spatial Josephson problem and the internal Josephson problem is that in the latter atoms in different internal states interact, while in the former two particles in different reservoirs do not. However if the interaction between the two internal states is taken to be zero
\begin{equation}
\lambda_{12}=0,
\end{equation}
then equations for the internal Josephson problem reduce to that of the spatial problem \cite{zsl98,rsf99,sfg97}. thus the internal effect displays all the phenomena the spatial Josephson effect has.

As in the spatial Josephson problem there are three different regimes depending on the strength of the coupling between the internal states \cite{leg01}. If the external field is much stronger than the interactions
\begin{equation}
|\vec{V}| \gg \lambda n,
\end{equation}
the system is said to be in the Rabi regime. (Here $\lambda$ can be taken to be the largest of $\lambda_{\alpha \beta}$ and $n$ the total density.) In this regime interactions are not important and the system experiences large oscillations in the density of each internal state. In the opposite regime, when the external field is extremely weak,
\begin{equation}
|\vec{V}| \ll \lambda n,
\end{equation}
the system is in the Fock regime. The amplitude of  density oscillations is negligible and the phase difference between the two states evolves as in the free precession case discussed in the previous section. Between these two regimes is the Josephson regime, for which,
\begin{equation}
|\vec{V}| \sim \lambda n.
\end{equation}
 In this regime, the density imbalance between the two internal states go through small oscillations, similar to the usual spatial Josephson effect. 

All three regimes are described by Eq.[\ref{spineom}], which reduces to 
\begin{eqnarray}
    \dot{\vec{S}_c}  &=& \vec{S}_c  \times  \vec{B}_c +  2 \vec{S}_c  \times
    \vec{V} \\  \nonumber 
\vec{B}_c &=& [ (\lambda_{1  1} - \lambda_{2
    2})    \rho_0  +  (\lambda_{11}+\lambda_{22}- 2  \lambda_{12})  
    \vec{S}_c  \cdot \hat{z}] \hat{z},   
\end{eqnarray} 
at zero temperature.

As the dynamics  conserves the magnitude of the  spin we only have two   dynamical  variables,   which   can  be   taken  as   angles $(\theta,\phi)$  in the  spherical polar  coordinates representing the orientation  of the spin.  Another conserved  quantity in this dynamics is  the Hamiltonian introduced  earlier, 
\begin{eqnarray}
{\cal H}  =  (  \lambda_{11} -\lambda_{22})  \bar{\rho}_0  \hat{z} \cdot \vec{S}_c    + \frac{1}{2} J^{ij} S_c^i  S_c^j +  2 \vec{S}_c  \cdot \vec{V}.
\label{onespinh}
\end{eqnarray}

On  the sphere  defined by  $(\theta,\phi)$, contours  of constant ${\cal  H}$  will define  the  paths  along  which the  spin  will precess.   We can  write  the  Hamiltonian as  a  function  of $(\theta,\phi)$ as 
\begin{equation} 
{\cal  H} = A \cos^2(\theta) + B \cos  (\theta) + C \sin(\theta) \cos(\phi)  + D,  
\end{equation}
and identify  
\begin{eqnarray} 
A &=& \frac{1}{2}  ( \lambda_{11} + \lambda_{22} -  2 \lambda_{12}) |S_c|^2 \\ \nonumber  
B &=& \left( (\lambda_{11} -  \lambda_{22}) \bar{\rho}_{0} |S_c| +  2 V^z |S_c| \right)  \\  \nonumber  
C &=&  2  V^x  |S_c|  \\ \nonumber  
D  &=& \lambda_{12}  |S_c|^2. 
\end{eqnarray} 
    
At  strong external  fields  the paths  are  almost circular  with centers  on  the line  oriented  along  vector $\vec{V}$,  passing through the origin, as in  the usual Rabi problem, and significant changes in the  populations occur throughout the course  of a Rabi oscillation. As a function  of $\theta$ and $\phi$ the Hamiltonian has one maximum and one  minimum, thus all the trajectories circle these   extremum   points,  if   the   external  field   satisfies
 \begin{equation} 
\label{condition} 
\left|  \frac{C}{2 A} \right| > ( 1 - \left|\frac{B}{2 A}\right|^{2/3} )^{3/2}.
\end{equation}

When the intrinsic Rabi frequency $|\vec{V}|$ becomes the order of the  density  caused  shifts  $\sim  \lambda  n$,  We get into the Josephson regime. Even on  resonance, large  population transfers from one state to another does  not take place. This can be easily seen from the structure of trajectories on the Bloch sphere.  When the external field does  not satisfy (\ref{condition}), instead of having  one  minimum and  one  maximum,  the  Hamiltonian has  two maxima,  one minimum  and  a  saddle point.   The  two constant  H trajectories crossing at the saddle point separate the sphere into three regions, giving us  three kind of trajectories, circling around the minimum  or one  of the  two  maxima. None  of these  oscillations however,  result  in  large  population  transfers,  which  is  in contrast with  the Rabi problem, in which one can  change internal states  of  all the  atoms  with  an  arbitrarily small  field  on resonance. Josephson  effect can be  understood if we realize  that the transition frequency for an atom depends on the populations. So for weak  fields   even  a  small  population   transfer  carries  the transition  away from  resonance, {\it  i.e.} makes  the effective detuning much larger than the Rabi frequency.

We can  calculate the oscillation  frequency along each path, for both high and low fields. The period of the precession along a path ${\cal C}$ on which   ${\cal  H}(\theta,\phi)   =   {\cal  H}'$   is  given   by
 \begin{equation} 
T({\cal H}')  = \int_{\cal C} dl \frac{1}{|\nabla {\cal  H}|}.   
\end{equation}  
By  converting the  integral  to  a surface integral over a $\delta$ function and integrating over the angle  $\phi$,  we  have  
\begin{equation}  
\label{integral}  
T  = \frac{2}{|A|} \int dx \frac{1}{\sqrt{x^4 + C_3 x^3 + C_2 x^2 + C_1 x  +  C_0}}  
\end{equation}   
where  
\begin{eqnarray}  
C_3  &=&  2  \frac{B}{A}   \\  \nonumber  
C_2   &=&  \left(\frac{C^2+B^2}{A^2}- 2\frac{H-D}{A}\right) \\ \nonumber  
C_1 &=& -2 \frac{B (H-D)}{A^2} \\ \nonumber 
C_0 &=& - \frac{C^2 - (H-D)^2}{A^2}.  
\end{eqnarray}  

In  the  high  field  case,  for  the  external  field  satisfying (\ref{condition}),  the  Hamiltonian   will  take  values  between $H_{max}$  and $H_{min}$,  the values  of the  Hamiltonian  at the maximum and  minimum points.   Any value of  H in this  range will uniquely correspond to one trajectory  and the period of motion on such a trajectory will  be given by 
\begin{eqnarray} 
\label{imagK} 
T&=&\frac{4}{|A|  \sqrt{p  q}}  K(\frac{1}{2}  \sqrt{\frac{(x_1  - x_2)^2  - (p-q)^2}{p q}})  \\ \nonumber  
p^2&=&(m-x_1)^2 +  n^2 \\ \nonumber  
q^2&=&(m-x_2)^2 +  n^2  
\end{eqnarray} 
where  $K$ is  the complete  elliptic integral  \cite{arfken}, $x_1$,  $x_2$  are the real roots, m is the real part  and n is the absolute value of the imaginary part of the remaining two complex conjugate roots of the polynomial in Eq.(\ref{integral}).

In  the  weak  field  case  there  are 4  special  values  of  the Hamiltonian, $H_{min}$, the  minimum value, $H_{saddle}$ the value at the saddle point, $H_{max1}$ and $H_{max2}$, the smaller and the larger of  the values  at the  two maxima respectively (See  Figure 1).   For $H$ values in the range $H_{max2} >  H' > H_{max1}$ or $H_{saddle} > H > H_{min}$  there is  again one  to one  correspondence  between H values and  trajectories on the sphere. For  such trajectories the frequencies are again given by  (\ref{imagK}). For the values of $H$ satisfying $ H_{max1} > H > H_{saddle}$ there are two trajectories corresponding to each $H$, one  circling around $ H_{max1} $, and the other circling around $H_{max2}$. However they  both have the same period given by  
\begin{equation}  
T=\frac{4}{|A| \sqrt{  (x_4  -  x_2) (x_3  - x_1)}}  K(\sqrt{\frac{(x_2-x_1)  (x_4-x_3)}{(x_4  -  x_2)  (x_3  - x_1)}})  
\label{realK} 
 \end{equation} 
where  $K$  is again  the complete  elliptic integral and $x_1 <  x_2 < x_3 < x_4 $ are  the four real roots of the polynomial in Eq.(\ref{integral}).

A typical plot of frequencies for  the weak field case is given in figure  2.    Near  the   saddle  point  trajectories   slow  down logarithmically   in  $H-H_{saddle}$  as   expected  from   a  two dimensional dynamics.

We can  see for $\lambda_{11} = \lambda_{22}$ that our equations reduce  to  those  obtained  in \cite{wwc99,rsf99}  by  using  two coupled Gross-Pitaevskii  equations.  At high field  we get almost circular  trajectories,  and  correspondingly  large  oscillations between  internal  levels  of  the  condensate,  as  in  the  Rabi problem. For the low field  case we get three kind of trajectories all of  which give little  population change, these  correspond to Josephson oscillations.  Two of  these three kinds complete a full cycle around  the z axis, while  the third is trapped  in a region for  which $  \phi_{min} <  \phi  < \phi_{max}  $. Recalling  that $\phi$ represents the relative phase of the two condensates, we see that  these  trajectories  correspond  to  Josephson  oscillations between the  two internal states, caused by  the weak link  of the external  field.   The  other  two  kinds  of  trajectories  again correspond to  Josephson oscillations,  however in these  class of Josephson oscillations  there is  a $2 \pi$  phase slip  for every period of population change. such oscillations are called $\pi$ Josephson oscillations \cite{sfg97}.

It has been shown for the spatial Josephson effect that the system can dynamically maintain a population imbalance between the two macroscopically occupied states \cite{sfg97}. We can ask whether the internal Josephson effect also shows this ``macroscopic quantum self trapping'' effect. This question is meaningful only if a symmetry exists between states $1$ and $2$
\begin{equation}
\lambda_{11}=\lambda_{22}.
\end{equation}
If the internal states have different interactions a population imbalance is to be expected in any oscillation as the system will be biased to choose the internal state with lower interaction energy.

If we take $\lambda_{11}=\lambda_{22}$, our condition for the appearance of the extra maximum in the Hamiltonian becomes
\begin{equation}
|C| > 2 |A|,
\end{equation}
as $B=0$ in this case. One can immediately see that out of the three types of trajectories discussed for the general case, those corresponding to $\pi$ Josephson oscillations become macroscopically self trapped oscillations for $\lambda_{11}=\lambda_{22}$.

It is remarkable that none of the arguments given for the internal Josephson effect depend on the macroscopic occupation of a spatial state, {\it i.e.} Bose condensation. In fact if we assume that the system is not condensed
\begin{equation}
T > T_c,
\end{equation}
we get the equation for the precession of the normal spin from Eq.(\ref{spineom})
\begin{eqnarray}
\dot{\vec{S}}  &=& \vec{S}  \times  \vec{B}_n +  2 \vec{S}  \times \vec{V} \\  \nonumber 
\vec{B}_n &=& [ (\lambda_{1  1} - \lambda_{2
    2})  2  \rho_0  +  (\lambda_{11}+\lambda_{22}- 2  \lambda_{12})  2 \vec{S}  \cdot \hat{z}] \hat{z}.   
\end{eqnarray}

This equation is very similar to the equation for the condensate spin, and all the results found in this section apply for the non-condensed case if we identify
\begin{eqnarray}
 A &=&  (  \lambda_{11}  +
    \lambda_{22} - 2  \lambda_{12}) |S|^2 \\ \nonumber B  &=& \left( 2
    (\lambda_{11} - \lambda_{22}) \rho_{0} |S|  + 2 V^z |S| \right) \\
    \nonumber  C &=&  2  V^x |S|  \hspace{1cm}  \\ \nonumber  D &=&  2
    \lambda_{12} |S|^2.  \end{eqnarray}
The behavior of the fully condensed and non-condensed cases are similar, however the values of the precession frequencies and critical value of the external field are different. This difference, again, is a result of the fact that there are no exchange scattering processes in a Bose condensate.

From the  above discussion we come to the  conclusion that to observe the internal  Josephson effect it is not  required to have two Bose condensed samples. Two  words of caution should be voiced about this observation. First we  have only considered the coherent collisions and  the coherence of  the phase  of two  internal levels  will be destroyed on  a time scale  that is set  by mean free time  in the gas. Any  observed internal Josephson oscillation in a non-condensed sample  should decay in this time scale.  Second, to be  able to see this effect one has to go  to   very  low  field  strengths  such   that  the  population oscillations should be observable in a non-condensed sample.

Still we  have shown that it  is not absolutely  necessary to have Bose condensation  to observe  small oscillations in  the relative phase of the  internal states under a small  excitation field. Let us imagine a  non-condensed gas of atoms put  into a superposition of two  internal states by  a $\pi /  2$ pulse. If this  sample is further  subjected to  weak mixing  field on  resonance  one would naively expect population transfer  from one internal level to the other with the Rabi frequency of the field. However our discussion shows that  if the sample  and the field  satisfy 
\begin{equation} 
\tau_{\rm coll}^{-1}  = 8  \pi a_s^2 n  v_{\rm T} \ll  |\vec{V}| \ll \lambda n
\end{equation} 
we would have small  oscillations of the phase and  populations, which is  exactly what is observed  in the internal Josephson effect with condensed samples.

\section{Absorption resonances for a partially condensed gas: The Two spin Problem}
After analyzing  the fully  condensed and non-condensed  Bose gas, both of which can be represented by only one spin in our equations (\ref{spineom}), we  turn our attention to  the partially condensed Bose gas.   When the temperature is between  zero and condensation temperature,  $T_{\rm  BEC}$, there  is  both  a  condensed and  a non-condensed density  present. If these  densities are comparable we  have to  use the  full form  of Eq.(\ref{spineom}),  with both spins present.

We first  start with  the free precession  problem, {\it  i.e.} by setting the external field equal to zero. In this case we have the equations 
\begin{eqnarray} 
\label{freeS} 
\dot{\vec{S}} &=& \vec{S} \times  \vec{B}_n + 2  \lambda_{1 2}  \vec{S} \times  \vec{S}_c \\ 
\label{freeSc} \dot{\vec{S}_c} &=&  \vec{S}_c \times \vec{B}_c + 2  \lambda_{1  2}  \vec{S}_c   \times  \vec{S}.   
\end{eqnarray}  
The effective magnetic  fields $\vec{B}_n$ and  $\vec{B}_c$ both point in    the    $\hat{z}$    direction    and    are    defined    in Eq.(\ref{magfields}). As their  time derivatives are perpendicular to the  spin vectors, norm of  both of the  spins, $|\vec{S}|$ and  $|\vec{S}_c|$ are  conserved. This conservation  simply means that the  total  number  of  atoms  in  the  condensate  and  over  the condensate  are  conserved.   Another  conserved quantity  can  be obtained  by  adding  Eq.(\ref{freeS})  to  Eq.(\ref{freeSc})  and taking   the  $\hat{z}$   component.   We  have   
\begin{equation} 
\frac{d}{dt}  \left(  \hat{z}  \cdot  (\vec{S}+\vec{S}_c)  \right) \equiv \frac{d}{dt} S^z_{\rm  tot} = 0.  
\end{equation} 
Physically this conservation law corresponds to  the fact that in the absence of a coupling field the total number of atoms in internal states 1 and 2 are conserved  separately. Although these three conservation laws  restrict  the resulting  dynamics  considerably, they  still allow  oscillations   in  which  the  density   of  condensed  and non-condensed atoms  in internal state  1 change, while  state two goes through  the same oscillations out  of phase with  1 to leave the total  number of  condensed and non-condensed  atoms constant. In our  spin representation the  degree of freedom  that expresses these oscillations will be  $S^z$, or equivalently $S^z_c$ as they add up to a constant.

The  other  conserved quantity  is  the Hamiltonian  (\ref{spinH}) which we  choose to rewrite  in terms of the  conserved quantities $S^z_{\rm tot}$ , $|S|$ , $|S_c|$ and the dynamical variable $S^z$ as 
\begin{eqnarray} 
\label{twospinH}  
{\cal H} &=& (\lambda_{11} - \lambda_{22}) (\rho_0 +  \bar{\rho}_0) S^z_{\rm tot} + \frac{1}{2} (\lambda_{11} +  \lambda_{22} - 2  \lambda_{12}) (S^z_{\rm tot})^2 \\  \nonumber 
&+& 2  \lambda_{12} |S|^2  + \lambda_{12}  |S_c|^2 + (\lambda_{11}  -   \lambda_{22})  \rho_0  S^z   \\  \nonumber  
&+& \frac{1}{2} (\lambda_{11} + \lambda_{22} - 2 \lambda_{12}) (S^z)^2 + 2 \lambda_{12} \vec{S} \cdot \vec{S_c}.  
\end{eqnarray}

To  investigate   the  oscillations  of  the   degree  of  freedom physically described  above and represented  by $S^z$ we  take the equation  of motion  for $S^z$  
\begin{equation}  
\frac{d S^z}{dt} =\hat{z}  \cdot \dot{\vec{S}}  =  2 \lambda_{12}  \hat{z} \cdot  ( \vec{S}   \times  \vec{S}_c   )   =  2   \lambda_{12}  {\cal   V}. 
\end{equation}  
We define  ${\cal  V}$  to be  the  volume of  the parallelepiped formed  by the vectors  $\hat{z}$ , $  \vec{S}$ and $\vec{S}_c$.  We can express the  absolute value of this volume in terms   of  the  inner   products  of   these  three   vectors  as
 \begin{equation}  
|{\cal V}|  =  \sqrt{ |S|^2  |S_c|^2 -  (\vec{S} \cdot \vec{S}_c)^2  - |S_c|^2  (S^z)^2 - |S|^2  (S_c^z)^2 +  2 S^z S_c^z \vec{S} \cdot \vec{S}_c}.  
\end{equation}

Now  we  can solve  for  $\vec{S}  \cdot  \vec{S}_c$ in  terms  of conserved quantities and $S^z$ from Eq.(\ref{twospinH}). This will give us the  equation of motion for $S^z$  expressed only in terms of  conserved   quantities  and  $S^z$   itself:  
\begin{equation} 
\label{polyeqn}  
\left|  \frac{d   S^z}{dt}  \right|  =  \sqrt{C_4 (S^z)^4   +C_3  (S^z)^3   +C_2  (S^z)^2   +C_1  (S^z)   +   C_0  }
 \end{equation} 
The value of $S^z$ will oscillate between $x_1$ and $x_2$ which are two roots of the polynomial inside the square root in Eq.(\ref{polyeqn}). For values of $S^z$ in the interval $ x_1 < S^z  < x_2$,  this polynomial  takes positive  values.  When $S^z$ reaches  its maximum  or minimum  value, the  vectors  $\vec{S}$ , $\vec{S}_c$  and  $\hat{z}$  are   coplanar.  This  allows  us  to integrate  Eq.(\ref{polyeqn})  without   paying  attention  to  the absolute value. We can express  the period of $S^z$ as an integral of the form in Eq.(\ref{integral})
\begin{equation}
T_z  =  2 \int_{x_1}^{x_2}  d  S^z  \frac{1}{\sqrt{  C_4 (S^z)^4  +C_3 (S^z)^3 +C_2 (S^z)^2 +C_1 S^z +C_0 }}.
\end{equation}

Due to the  abundance of conserved quantities in  the two spin problem, the expressions for the coefficients are more complicated compared to the one spin case 
\begin{eqnarray} 
C_4 = &-& \frac{\Delta_2}{2} ( \frac{\Delta_2}{2} -  4 \lambda_{12} ) \\ \nonumber  
C_3 = &-& [ \Delta_1  \Delta_2 \rho_0  + 4  \lambda_{12}  ( \frac{\Delta_2}{2} S_{\rm tot}^z - \Delta_1 \rho_0)]  \\ \nonumber 
C_2 = &[& \Delta_2 ( {\cal  H} -  \Delta_1 (\rho_0 +  \bar{\rho}_0 ) S_{\rm  tot}^z - \frac{\Delta_2}{2}  (S_{\rm  tot}^z)^2 -  2  \lambda_{12} |S|^2  + \lambda_{12} |S_c|^2) \\ \nonumber 
&-& 4 \lambda_{12} ( {\cal H} - \Delta_1 \bar{\rho}_0  S_{\rm tot}^z -  \frac{\Delta_2}{2} (S_{\rm tot}^z)^2 + \lambda_{12} |S|^2)] \\  \nonumber 
C_1 = &[& ({\cal H} - \Delta_1   (\rho_0   +    \bar{\rho}_0   )   S_{\rm   tot}^z   - \frac{\Delta_2}{2}  (S_{\rm  tot}^z)^2 -  2  \lambda_{12} |S|^2  + \lambda_{12} |S_c|^2) (2 \Delta_1 \rho_0 + 4 \lambda_{12} S^z_{\rm tot}) + 8 \lambda_{12}^2 |S|^2  S_{\rm tot}^z ] \\ \nonumber 
C_0 = &-& ({\cal H} - \Delta_1  (\rho_0 + \bar{\rho}_0 ) S_{\rm tot}^z - \frac{\Delta_2}{2}  (S_{\rm  tot}^z)^2 -  2  \lambda_{12} |S|^2  + \lambda_{12} |S_c|^2)^2  \\ \nonumber 
&+&  2 ({\cal H}  - \Delta_1 (\rho_0 + \bar{\rho}_0  ) S_{\rm tot}^z) ( 2  \lambda_{12} |S|^2 + \lambda_{12} |S_c|^2)  - 4  \lambda_{12}^2 |S|^4 \\  \nonumber
&-& \lambda_{12}^2  |S_c|^4 -  \Delta_2  \lambda_{12} |S_c|^2  (S_{\rm tot}^z)^2  -  2 (\Delta_2  +  2  \lambda_{12}) \lambda_{12}  |S|^2 (S_{\rm    tot}^z)^2,    
\end{eqnarray}    
with    the    notation 
\begin{eqnarray}  
\Delta_1 &=&  (\lambda_{11}  - \lambda_{22})  \\ \nonumber   
\Delta_2  &=&   (\lambda_{11}  +   \lambda_{22}   -  2 \lambda_{12}).  
\end{eqnarray}

As in  the one  spin case this  integral can be  exactly evaluated \cite{GR}.  If all  the roots of the polynomial  inside the square root in Eq.(\ref{polyeqn}) are real we have 
\begin{equation} 
T_z = \frac{4}{\sqrt{C_4}}  \frac{1}{\sqrt{(x_4 - x_2)  (x_3 -  x_1)}} K \left( \sqrt{  \frac{(x_2 - x_1) (x_4  - x_3)}{(x_4 -  x_2) (x_3 - x_1)}} \right).  \end{equation} 
Here $x_3 < x_4$ are the remaining real root of the polynomial in equation of motion (\ref{polyeqn}), which are assumed  to be real. In the case  of imaginary $x_3$ and $x_4$, an  analogue of  Eq.(\ref{imagK}) will give  the expression for the period.

The  second oscillation  for  free precession  corresponds to  the precession of  total phase  about the z  axis. This  precession is affected  by  the  population  oscillations found  above, and  the oscillation frequency is not as easily calculated. We can generally describe its motion as a  sum of two components. One corresponding to a uniform  precession around the z axis  with the density shift as  in Eq(\ref{densshift}),  and  the other  corresponding to  the effect of  an oscillating  magnetic field in  $\hat{z}$ direction, caused  by  the   population  oscillations  discussed  above.  The coupling between these  two components is best seen  when we write the  equation  of  motion  for  the spin  components  in  the  $x-y$ plane. Defining $S^+ = S^x + i S^y $ and $ S_c^+= S^x_c + i S^y_c$ we have
\begin{eqnarray}
\label{roteqn}
i  \dot{S}^+  &=&   (B_n(t)  +  2  \lambda_{12}  S^z_c(t))   S^+  -  2 \lambda_{12}  S^z(t)  S^+_c  \\   \nonumber  
i \dot{S}^+_c  &=&  -  2 \lambda_{12} S^z_c(t) S^+ +(B_c + 2 \lambda_{12} S^z(t)) S^+_c.
\end{eqnarray}

We have  seen that  when $S^z$ reaches  its maximum or  minimum values $\vec{S}$  and $\vec{S}_c$  are in  the  same vertical  plane. We  can calculate exactly  how much the spins  have rotated around  the z axis throughout the  course of one  $S^z$ oscillation. The effect  of $S^z$ oscillations will present itself through the integral
\begin{equation}
I =  \int_0^{T_z} dt  S^z(t) = 2  \int_{x_1}^{x_2} dS^z \frac{S^z}{|\frac{d
S^z}{dt}|}
\end{equation}
which is again exactly calculable.

To calculate the rotation  angle we integrate Eq.(\ref{roteqn}). After time $T_z$, $S^+$ and $S^+_c$ will be given by
\begin{equation}
\left[ \begin{array}{c}  S^+(T_z) \\ S^+_c(T_z)  \end{array} \right] = e^{i {\cal M}} \left[  \begin{array}{c} S^+(0) \\ S^+_c(0) \end{array} \right]
\end{equation}
where ${\cal M}$ is a two by two matrix with elements
\begin{eqnarray}
{\cal M}_{11} &=& \Delta_1 (2  \rho_0 + \bar{\rho}_0) T_z + ( \Delta_2+ 2 \lambda_{12})  S^z_{\rm tot} T_z + (\Delta_2 -  2 \lambda_12) I \\ \nonumber  
{\cal M}_{12}  &=&  -2 \lambda_{12}  I  \\ \nonumber  
{\cal M}_{21} &=&-2  \lambda_{12} S^z_{\rm  tot} T_z +  2 \lambda_{12}  I \\ \nonumber 
{\cal  M}_{22} &=& \Delta_1  (\rho_0 + \bar{\rho}_0 )  T_z + \Delta_2 S^z_{\rm tot} T_z + 2 \lambda_{12} I.
\end{eqnarray}

Instead of giving the resulting long expression for the rotation angle in  one  period,  we  choose  to  describe  the  motion  qualitatively. The precession of the two spins are affected by the competition between two effects. The first effect is,  due  to the absence of exchange scattering in the condensate, the condensate spin $\vec{S}_c$ sees an effective  magnetic field $B_c$ different from  the  effective magnetic  field seen by  the  normal gas  spin $\vec{S}$.  The   second,  as  discussed  above,   is  the  condensate population oscillations characterized by $S^z$.

If there is not much  difference between the densities of two internal states,  both spins lie  close to  the x-y  plane, and  their relative phase oscillates  around zero, never growing large.  However, if there is a lot of density  difference between two internal states, the spins are close to the $\hat{z}$ axis. Over one period of $S^z$ oscillation, the phase difference can be a multiple of $2 \pi$.

We can investigate the precession easily in the limit when both of the spins are almost aligned with the $\hat{z}$ axis. We can write linear equations for the perpendicular components of the spins and get     the     two    precession     frequencies:
\begin{equation} 
\omega=\tilde{\omega} + ( w_2 - w_1)  + (\lambda_{12} (n_1 - n_2) + 2 \lambda_{2 2}  n_2  - 2 \lambda_{1 1}  n_1).  
\end{equation}
Where $\tilde{\omega}$ satisfies, 
\begin{equation} 
(\tilde{\omega} +  \lambda_{1 2}  \Delta  n_{t}) (\tilde{\omega}  - \lambda_{1  1} n_{c1} + \lambda_{2 2} n_{c2}) = \lambda^2_{1 2} \Delta n_c \Delta n_t.  
\end{equation} 
with  
\begin{equation} 
\Delta  n_{(c,t)} =  n_{2  (c,t)} -  n_{1 (c,t)}.   
\end{equation}

If  the mixing  angle is  not small  one would  expect to  see two different  frequencies  in  general.    The  average  of  the  two frequencies will  be controlled by the average  density shift seen  by an  atom in  the sample, while  the splitting will  reflect the average rate the condensate fraction of one of the internal states oscillates \cite{ole99}.

The appearance  of a second  frequency should be detectable  in an experiment  that  probes  a  partially  condensed Bose  gas  in  a superposition state. We propose using a partially condensed gas in a Ramsey  separated field arrangement,  as in the  fountain atomic clocks \cite{gch93,lgi98}. In this case the appearance of a second frequency  would  present  itself  as  a  beating  in  the  Ramsey fringes. This  beating however, will  vanish both in the  limit of full condensation  and the  limit of a  normal gas, and  should be most  prominent  when  the  condensate  fraction  is  close  to  a half. The exact values of  the frequencies can be obtained solving the equations of motion numerically.

Here we want to remind the  reader that the equations used in this section Eq.(\ref{spineom})  were derived for a  uniform system. If the particles are cold enough  and the condensate is prepared in a shallow trap  to make sure that  the movement of each  part of the cloud is negligible in the  center of mass coordinate frame during the time of measurement, the equations will be locally correct and the  experiment  should show  a  density  averaged  result in  the precession  frequencies. Otherwise  the  effects of  inhomogeneity must   be   included    using   the   full   transport   equations Eq.(\ref{transport},\ref{continuity}).

When an  external field  is turned on  the $\hat{z}$  component of total spin  in not  conserved anymore, there  is a  net population transfer  from one  internal state  to the  other.  Much  like the single spin  case we  have two limits.   When $|\vec{V}|$  is much larger  then the  density shift  $\lambda  n$ we  have both  spins following almost  circular trajectories around  $\vec{V}$. For the weak  field  case  the  system   can  be  best  described  as  two non--linear oscillators going through coupled oscillations with an occasional  two $\pi$ phase  slip for  one of  them. In  this most general  case precession  frequencies  can be  found by  numerical integration of Eq.(\ref{spineom}).

\section{Fermions}

Finally to understand the  effect of statistics better we consider the  same  problem  for  a  Fermi  gas.  We  will  have  the  same Hamiltonian  Eq.(\ref{spatialH}),  however,  the  Fermionic  field  operators       will       satisfy 
\begin{equation}       
\{\psi_{\alpha}(r),\psi^+_{\beta}(r')  \}  =  \delta_{\alpha  \beta} \delta(r-r'),    
\end{equation}   
where   $\{,\}$    denotes   the anti--commutator. The derivation will process along the same lines with the Bose case. The effect of statistics will be seen whenever we  average  a  four  particle  operator.  The  exchange  term  in Eq.(\ref{wick})     will     change     sign:     
\begin{equation}
 \label{wickfermi}     
\langle    \psi^+_{\alpha}    \psi^+_{\beta} \psi_{\beta}  \psi_{\alpha}   \rangle  =  \langle  \psi^+_{\alpha} \psi_{\alpha} \rangle  \langle \psi^+_{\beta} \psi_{\beta} \rangle - \langle    \psi^+_{\alpha}    \psi_{\beta}    \rangle    \langle
  \psi^+_{\beta} \psi_{\alpha} \rangle.  
\end{equation}

As a result  we get the transport equation  for the density matrix defined  as in  Eq.(\ref{normalmatrix}) with  sign changes  in the terms    corresponding     to    the    exchange    contributions: 
\begin{eqnarray}       
\label{transportfermi}      
\lefteqn{\left( \partial_{\rm  t} + \frac{\vec{p}}{m}  \cdot \nabla_r  - \nabla_r( \frac{U_{\alpha}(r)+U_{\beta}(r)}{2}  )   \cdot  \nabla_p  \right) \varrho_{\alpha  \beta}  (r,p)  =}   \\  \nonumber  
&+&  i  \left[ U_{\alpha}(r)   -U_{\beta}(r)  +   \sum_{\gamma}  (\lambda_{\gamma \alpha} - \lambda_{\gamma  \beta}) \rho_{\gamma \gamma}(r) \right] \varrho_{\alpha  \beta}  (r,p) \\  \nonumber  
&-& i  \sum_{\gamma} \lambda_{\gamma  \alpha}  \rho_{\alpha \gamma}(r)  \varrho_{\gamma \beta} (r,p) + i \sum_{\gamma} \lambda_{\beta \gamma} \rho_{\gamma \beta}(r)   \varrho_{\alpha  \gamma}   (r,p)   \\  \nonumber   
&+& \sum_{\gamma}   \frac{\lambda_{\gamma  \alpha}   +  \lambda_{\beta \gamma}}{2}   \nabla_r  \rho_{\gamma  \gamma}(r)   \cdot  \nabla_p \varrho_{\alpha  \beta}  (r,p)   \\  \nonumber  
&-&  \sum_{\gamma}  \lambda_{\gamma  \alpha}  \nabla_r  \rho_{\alpha \gamma}(r)  \cdot \nabla_p    \varrho_{\gamma   \beta}    (r,p)    +   \sum_{\gamma} \lambda_{\beta  \gamma}   \nabla_r  \rho_{\gamma  \beta}(r)  \cdot \nabla_p  \varrho_{\alpha   \gamma}  (r,p)  \\   \nonumber  
&+&  i \sum_{\gamma}   \left[   V_{\gamma  \alpha}(r,t)   \varrho_{\gamma \beta}(r,p)  - V_{\beta \gamma}(r,t)  \varrho_{\alpha \gamma}(r,p) \right] \\ \nonumber  
&+& \frac{1}{2}\sum_{\gamma} \left[ \nabla_r V_{\gamma \alpha}(r,t)  \cdot \nabla_p \varrho_{\gamma \beta}(r,p) - \nabla_r  V_{\beta \gamma}(r,t)  \cdot  \nabla_p \varrho_{\alpha \gamma}(r,p) \right].  
\end{eqnarray}

From the  transport equation by assuming all  the interactions and the sample to be spatially  homogeneous we can get the equation of motion  for  the internal  state  density matrix  
\begin{eqnarray}  
\dot{\rho}_{\gamma \gamma'}= &i& (w_{\gamma}-w_{\gamma'}+U_{\gamma}-U_{\gamma'}) \rho_{\gamma \gamma'} -  i \sum_{\alpha} (\lambda_{\alpha \gamma}-\lambda_{\alpha \gamma'}) \rho_{\gamma \alpha} \rho_{\alpha \gamma'} \\  \nonumber 
+&i& \sum_{\alpha} ( V_{\alpha \gamma}(t)   \rho_{\alpha   \gamma'}   -   V_{\gamma'   \alpha}(t) \rho_{\gamma \alpha}) 
\end{eqnarray} 
with $U_{\gamma}$ are defined as,  
\begin{equation}   
U_{\alpha}=  \sum_{\beta}  \lambda_{\alpha \beta} n_{\beta}.  
\end{equation}

If we assume that only states 1  and 2 are coupled and all the off diagonal elements involving the other states are equal to zero, we get a very  simple dynamics.  The time derivative  of the diagonal elements do  not depend on  $\rho$ while the off  diagonal element $\rho_{12}$ changes  according to 
\begin{eqnarray} 
\label{fermi12}
\dot{\rho}_{12} =  &i& \left(  w_1 - w_2  + \sum_{\beta  \neq 1,2} (\lambda_{\beta   1}  -   \lambda_{\beta  2})   n_{\beta}  \right) \rho_{12}  \\  \nonumber
+&i&  \sum_{\alpha}  (  V_{\alpha  1}(t) \rho_{\alpha 2} - V_{2 \alpha}(t) \rho_{1 \alpha}) 
\end{eqnarray}

When  we  go to  the  Bloch  sphere  representation in  the  basis rotating   with    the   frequency   of    the   external   field: 
\begin{equation} 
\rho'_{\gamma  \gamma'} = \rho_{0} \delta_{\gamma \gamma'}   +   \vec{S_f}   \cdot   \vec{\sigma}_{\gamma   \gamma'}
 \end{equation}   
we   get    the   simple   equation   of   motion
\begin{equation}  
\dot{\vec{S}}_f=  2  \vec{S}_f  \times  \vec{V}.
\end{equation}  
Which  can   be  derived  from  the  corresponding Hamiltonian  
\begin{equation} 
\label{spinfermiH}  
{\cal  H}_f =  2
 \vec{S}_f \cdot \vec{V}.  
\end{equation}

In  the  mean field  picture  there  are  no coherent  effects  of interactions for  a transition  between two states.   The exchange contributions  to  the  precession  frequency exactly  cancel  the direct  contributions. For  short--range potentials,  the exchange contribution  to the  energy has  the same  absolute value  as the direct  contribution. For  Bosons  the two  contributions add  up, while for fermions they cancel.

The precession frequency for a free Fermi gas can be read off from Eq.(\ref{fermi12})   
\begin{equation}  
\omega=   (w_2  -   w_1)  + \sum_{\gamma\not=1,2}  (\lambda_{2 \gamma}  -  \lambda_{1 \gamma}) n_{\gamma}.  
\end{equation} 
This expression shows that the density dependent  frequency  shift  encountered  in the  fountain  atomic clocks can be eliminated if  a fermionic sample is used instead of a Bose  gas. Any contributions to  the frequency shift  will be of higher order  in the diluteness parameter  $a_s/(n^{-1/3})$ of the gas. Finally we  see that the behavior under  an external filed is not at all  different from the usual Rabi  precession. An analogue of internal Josephson effect does  not appear in this case since the ``energy'' of the system Eq.(\ref{spinfermiH}) does not depend on the density at all (there are no quadratic terms in the Hamiltonian in the spin representation).

    \section{Conclusion}

We studied the effect of  external interactions on the  internal dynamics  of atoms  in  a dilute gas. For a Bose gas
we derived a general transport equation valid for any partially condensed and/or non-uniform gas. We then applied it to the case of a homogeneous gas and investigated the effects of interactions on the internal degrees of freedom. As a first result we obtained an expression for the density induced frequency shift in atomic clocks, for a gas which is above BEC, or is at zero temperature. Furthermore we found that if a partially condensed sample is used in an atomic clock, one would get two density dependent frequencies instead of one, due to the exchange of atoms between the normal part and the condensed part of the gas. 

We then went on to analyze the effect of an external field. We show how Rabi oscillations are replaced by internal Josephson oscillations as the strength of the external field is reduced. We calculated the frequencies of both oscillations exactly. We have also found that an analogue of the internal Josephson effect should be observable for a non-condensed sample. 

Finally we considered a Fermi gas, and derived the transport equation. We demonstrated that it is possible to eliminate density shift in Rabi frequency by using a Fermi gas in an atomic clock and that no analogue of the internal Josephson effect is possible for a Fermi gas. 

M.\"{O}.O. is grateful to S.R. Shenoy, F. Sols and D. Stroud for useful discussions.

$^{\dagger}$ Current address: Department of Physics, The Ohio State University, Columbus OH 43210.   
\newpage

\begin{figure}
\epsfysize=7cm
\epsfxsize=7cm
\epsfbox{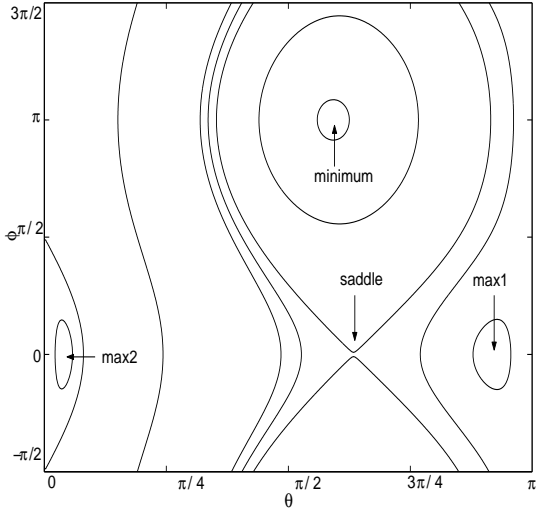}
\caption[FIG1.]{
\label{fig1}
The contour lines for the Hamiltonian  Eq.(\ref{onespinh}), on the Bloch sphere, defined by $\theta$ , $\phi$, for weak external fields. As the field strength is increased, the saddle point comes closer to one of the maxima and they destroy each other when the condition in Eq.(\ref{condition}) is satisfied, leaving just one maximum and one minimum. In the figure the trajectories encircling the maxima correspond to Josephson oscillations with an average phase difference of $\pi$, while those encircling the minimum are the usual Josephson oscillations corresponding to small oscillations of phase difference. The frequencies of motion along these trajectories are given in fig2.
}
\end{figure}

\begin{figure}
\epsfysize=7cm
\epsfxsize=7cm
\epsfbox{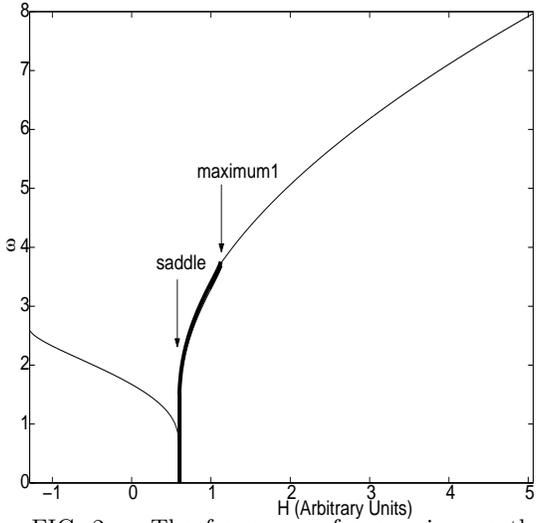}
\caption[FIG2.]{
\label{fig2}
The frequency of precession on the Bloch sphere as a function of the value the Hamiltonian Eq.(\ref{onespinh}) takes. Hamiltonian can take values from $H_{min}$ to $H_{max2}$. Near the saddle point the precession slows down logarithmically, as expected from a two dimensional dynamics. Between $H_{saddle}$ and $H_{max1}$ there are two trajectories for each value that the Hamiltonian takes. However, they both have the same frequency Eq.(\ref{realK}).}
\end{figure}


\newpage
    \begin{references}          

\bibitem{mhj98}    M.R~Matthews,    D.S.~Hall,    D.S.~Jin, J.R.~Ensher,  C.E.~Wiemann,  E.A.~Cornell, F.~Dalfovo,  C.~Minniti and  S.~Stringari,   Phys.  Rev.   Lett.  {\bf  81},   243  (1998)

\bibitem{hmw98} D.S.~Hall, M.R.~Matthews, C.E.~Wieman and E.A.~Cornell, Phys. Rev. Lett. {\bf 81}, 1543 (1998)

\bibitem{hme98} D.S.~Hall, M.R.~Matthews, J.R.~Ensher, C.E.~Wieman and E.A.~Cornell, Phys. Rev. Lett. {\bf 81}, 1539 (1998)

\bibitem{sac98} D.M.~Stamper-Kurn, M.R.~Andrews, A.P.~Chikkatur, S.~Inouye, H.J.~Miesner, J.~Stenger and W.~Ketterle, Phys. Rev. Lett. {bf 80}, 2027 (1998)

\bibitem{sis98} J.~Stenger, S.~Inouye, D.M.~Stamper-Kurn, H.J.~Meisner, A.P.~Chikkatur and W.~Ketterle, Nature {\bf 396}, 345 (1998)

\bibitem{mss99} H.J.~Meisner, D.M.~Stamper-Kurn, J.~Stenger, S.~Inouye, A.P.~Chikkatur and W.~Ketterle, Phys. Rev. Lett. {\bf 82}, 2228 (1999)

\bibitem{smc99} D.M.~Stamper-Kurn, H.J.~Miesner, A.P.~Chikkatur, S.~Inouye, J.~Stenger and W.~Ketterle, Phys. Rev. Lett. {\bf 83}, 661 (1999)

\bibitem{tlh98} T.L.~Ho, Phys. Rev. Lett. {\bf 81}, 742 (1998)

\bibitem{lpb98} C.K.~Law, H.~Pu and N.P.~Bigelow, Phys. Rev. Lett. {\bf 81}, 5257 (1998)

\bibitem{yip99} S.K.~Yip, Phys. Rev. Lett. {\bf 83}, 4677 (1999)

\bibitem{imo99} T.~Isoshima, K.~Machida and T.~Ohmi, Phys. Rev. A {\bf 60}, 4857 (1999)

\bibitem{kue00} M.~Koashi and M.~Ueda, Phys. Rev. Lett. {\bf 84}, 1066 (2000)

\bibitem{hyi00} T.L.~Ho and S.K.~Yip, Phys. Rev. Lett. {\bf 84}, 4031 (2000)

\bibitem{ued01} M.A.~Ueda, Phys. Rev. A {\bf 6301}, 3601 (2001)

\bibitem{kst01} U.~Al~Khawaja and H.~Stoof, Nature {\bf 411}, 918 (2001)

\bibitem{gch93} K.~Gibble and S.~Chu, Phys. Rev. Lett. {\bf 70},  1771    (1993)  

\bibitem{lgi98}    R.~Legere   and    K.~Gibble, Phys.   Rev.   Lett.  {\bf   81},   5780  (1998)   

\bibitem{krc89} M.A.~Kasevich, E.~Riis, S.~Chu and R.G.~DeVoe, Phys. Rev. Lett. {\bf 63} 612 (1989)

\bibitem{gll96}  S.~Ghezali,  Ph.~Laurent, S.~N.~Lea,  A.~Clairon, Europhys.Lett.{\bf  36},   25  (1996) 

\bibitem{morelevel}  For a  review of experiments where coherent effects between more than two levels are important see:
    M.O.~Scully  and   M.S.~Zubairy,  {\it  Quantum   Optics}  pp.245, Cambridge University Press, New York (1997 and references therein.

\bibitem{ldb01} C.~Liu, Z.~Dutton, C.H.~Behroozi and L.V.~Hau, Nature {\bf 409}, 490 (2001)

\bibitem{pfm01} D.F.~Phillips, A.~Fleischhauer, A.~Mair and R.L.~Walsworth and
M.D.~Lukin, Phys. Rev. Lett. {\bf 86}, 783

\bibitem{fkw98} D.G.~Fried, T.C.~Killian, L.~Willmann, D.~Landhius, S.C.~Moss, D.~Kleppner and T.J.~Greytak, Phys. Rev. Lett {bf 81}, 3811 (1998)

\bibitem{kfw98} T.C.~Killian, D.G.~Fried, L.~Willmann, D.~Landhius, S.C.~Moss, D.~Kleppner and T.J.~Greytak, Phys. Rev. Lett {bf 81}, 3807 (1998)

\bibitem{mah99a} M.R.~Matthews, B.P.~Anderson, P.C.~Haljan, D.S.~Hall, C.E.~Weimann and E.A.~Cornell, Phys. Rev. Lett. {\bf 83}, 2498 (1999)

\bibitem{ahr01} B.P.~Anderson, P.C.~Haljan, C.A.~Regal, D.L.~Feder, L.A.~Collins, C.W.~Clark and E.A.~Cornell, Phys. Rev. Lett. {\bf 86}, 2926 (2001)
 
\bibitem{mah99b} M.R.~Matthews, B.P.~Anderson, P.C.~Haljan, D.S.~Hall, M.J.~Holland, J.E.~Williams, C.E.~Weimann and E.A.~Cornell, Phys. Rev. Lett. {\bf 83}, 3358 (1999)

\bibitem{wbz99}    J.~Weiner,     V.S.~Bagnato,     S.~Zilio    and     P.S~Julienne, Rev. Mod.  Phys. {\bf 71}, 1  (1999) 

\bibitem{leg01} A.J.~Legget, Rev. Mod. Phys. {\bf 73}, 307 (2001)

\bibitem{bas81} E.P.~Bashkin, JETP  Lett. {\bf 33},  8 (1981);  Sov. Phys.  JETP {\bf  60}, 1122    (1985)

\bibitem{lla82} C.~Lhuillier and F.~Lalo\"{e}, Journal of Phys. (Paris) {\bf 43}, 197 (1982); Journal of Phys. (Paris) {\bf 43}, 225 (1982)

\bibitem{lru84} L.P.~L\'{e}vy and A.E.~Ruckenstein, Phys. Rev. Lett. {\bf 52}, 1512 (1984)

\bibitem{ole99} M.\"{O}.~Oktel and L.S.~Levitov, Phys. Rev. Lett. {\bf 83}, 6 (1999)

\bibitem{wwc99} J.~Williams,  R.~Walser, J.~Cooper, E.~Cornell and M.~Holland,  Phys. Rev.  A  {\bf 59},  R31 (1999)  

\bibitem{ost99} P.~{\"O}hberg and S.~Stenholm, Phys. Rev. A {\bf 59}, 3890 (1999)

\bibitem{vle99} P.~Villain and M.~Lewenstein, Phys. Rev. A {\bf 59}, 2250 (1999)

\bibitem{rwa97} J.~Ruostekoski and D.F.~Walls, Phys Rev. A {\bf 56}, 2996 (1997)

\bibitem{helium3} For a review of the internal Jospehson effect in He$^3$ see:
A.J.~Legget, Rev. Mod. Phys. {\bf 47}, 331 (1975)

\bibitem{jav86} J.~Javanainen, Phys. Rev. Lett. {\bf 57}, 3164 (1986)

\bibitem{zsl98}  I.~Zapata,  F.~Sols  and  A.J.~Leggett, Phys.  Rev.  A  {\bf  57}, R28  (1998)

\bibitem{rsf99} S.~Raghavan, A.~Smerzi,  S.~Fantoni and S.R.~Shenoy,  Phys. Rev. A {\bf 59}, 620  (1999)

\bibitem{sfg97}  A.~Smerzi, S.~Fantoni, S.~Giovanazzi and S.~R.~Shenoy, Phys. Rev. Lett. {\bf 79}, 4950 (1997)

\bibitem{kvg97} S.J.J.M.F.~Kokkelmans,  B.J.~Verhaar, K.~Gibble  and D.J.~Heinzen, Phys. Rev. A {\bf 56}, R4389 (1997)

\bibitem{vks87} B.~J.~Verhaar, J.~M.~V.~A.~Koelman,  H.~T.~C.~Stoof,  O.~J.~Luiten, Phys.Rev.{\bf A35},  3825  (1987)
  
\bibitem{tvs92}  E.~Tiesinga,  B.~J.~Verhaar, H.~T.~C.~Stoof,  D.~van~Bragt,  Phys.Rev.{\bf  A45},  2671  (1992)

\bibitem{API} C.~Cohen--Tannoudji,   J.~Dupont--Roc    and   G.~Grynberg,   {\it Atom-Photon  Interactions}   pp.  361,  Wiley,   New  York  (1998)

\bibitem{arfken}  G.B.~Arfken and H.J.~Weber,  {\it Mathematical  Methods for  Physicists}  pp. 333, Academic     Press,     San     Diego     (1995)     

\bibitem{GR} I.S.~Gradshteyn and I.M~Ryzhik, {\it Table of Integrals, Series and Products.}, fifth ed.  Integral number 3.145, pp 228, Academic Press, San Diego (1994)


    \end{references}
    \end{document}